\documentclass[12pt]{article}
\usepackage{a4wide,graphics,graphicx,amsmath,amssymb,cite,nicefrac}
\usepackage[verbose]{wrapfig}
\usepackage{authblk,hyperref}
\usepackage{url,amsfonts}
\usepackage[makeroom]{cancel}
\catcode`@=11 \@addtoreset{equation}{section} \catcode`@=12

\usepackage[english]{babel}
\usepackage{mathrsfs}
\usepackage{amsmath}
\usepackage{amsfonts}
\usepackage{amssymb}
\usepackage{makeidx}
\usepackage{graphicx}
\usepackage{physics}
\usepackage{float} %Include figure filesusepackage{graphicx} %Include figure files
\usepackage{subfloat}
\usepackage{multicol}
\usepackage{subcaption}
\begin{document}

\begin{titlepage}%1
\begin{center}
%\hfill DFPD/2017/TH/\\

\vskip 1.0cm

{\bf \Large Integrability and non-integrability for marginal deformations of 4d $\mathcal N = 2$ SCFTs}

\vskip 1.0cm

{ \large Jitendra Pal${}^{a^\star}$, Sourav Roychowdhury${}^{b^\dagger}$, Arindam Lala${}^{\clubsuit}$ and \\
Dibakar Roychowdhury${}^{a^\spadesuit}$ 
  }

\vskip 25pt

{\it ${}^a$%
Department of Physics, Indian Institute of
Technology
Roorkee,  \\ Roorkee
 247667, Uttarakhand, India}\\\vskip 5pt

 {\it ${}^b$School of Physical Sciences, Indian Association for Cultivation of Science, \\
Kolkata 700032, West Bengal, India}

\vskip 15pt
\texttt{%
${}^\star$jpal1@ph.iitr.ac.in},
\texttt{%
${}^\dagger$spssrc2727@iacs.res.in},
\texttt{%
${}^\clubsuit$arindam.physics1@gmail.com},
\texttt{%
${}^\spadesuit$dibakar.roychowdhury@ph.iitr.ac.in}

\end{%
center}

\vskip 1cm

\begin{%
center} {\bf Abstract}\\[3ex]\end{center}%
We study integrability and non-integrability for marginal deformations of 4d $\mathcal N =2$ SCFTs. 
We estimate various chaos indicators for the bulk theory which clearly shows the onset of a chaotic string dynamics in the limit of large deformations.
On the other hand, for small values of the deformation parameter, the resulting dynamics exhibits a non-chaotic motion and therefore presumably an underlying integrable structure.
Our analysis reveals that the $\gamma$-deformation in the type-IIA theory could be interpreted as an interpolation between a class of integrable $\mathcal N =2$ SCFTs and a class of non-integrable $\mathcal N =1$ SCFTs at strong coupling. 
We also generalise our results in the presence of the flavor branes.

%\today

%\end{center}%

%\noindent

%
\vfill

%\July 2008

\end{titlepage}

\newpage % \setcounter{page}{1} \numberwithin{equation}{section}
\tableofcontents

\section{Introduction and general idea of this paper}

Exploring the chaotic behaviour and the associated non-integrable structure in the context of gauge/gravity duality has been one of the thrust areas in modern theoretical physics
 for the last couple of decades.
In most of these examples,
 one finds chaotic motion of string trajectories \cite{PandoZayas:2010xpn,Basu:2011dg,Basu:2011di,Basu:2011fw,Basu:2012ae,PandoZayas:2012ig,Basu:2013uva,Basu:2016zkr,Panigrahi:2016zny,Giataganas:2013dha,Ishii:2021asw,Roychowdhury:2017vdo,Nunez:2018ags,Nunez:2018qcj,Filippas:2019puw,Banerjee:2018ifm,Rigatos:2020hlq}, while on the other hand, there are only handful of examples \cite{Sfetsos:2013wia,Hollowood:2014qma,Delduc:2013qra,Pal:2022hmh} that exhibit non-chaotic dynamics which therefore lead to an integrable structure.

Non-integrable systems are important in the context of the AdS/CFT correspondence \cite{Maldacena:1997re,Witten:1998qj}.
The key idea here is to analyse the classical phase space trajectory governed by the (semi)-classical string solutions that give rise to various chaos indicators. 
These indicators guarantee whether the associated phase space would be lacking the usual {\it{Kolmogorov-Arnold-Moser}} (KAM) tori characterising (quasi)-periodic orbits \cite{PandoZayas:2010xpn,Basu:2011dg,Basu:2011di}. 
Identification of these orbits is the key step towards unveiling an underlying integrable structure of the dynamical phase space at the classical level.

Following the holographic principle, it is conjectured that the semi-classical strings are dual to a class of single trace operators in the large $N$ limit.
Hence, the above framework leads to a conjecture about the integrability (or nonintegrability) of the strongly coupled dual superconformal field theory. 
There have been standard approaches to prove or disprove integrability of a classical dynamical system.
The most robust way to prove integrability is through the construction of Lax pairs \cite{Bena:2003wd,Arutyunov:2008if,Stefanski:2008ik,Sorokin:2010wn,Zarembo:2010sg}. 
Classical integrability of supergravity backgrounds containing an AdS$_5$ and AdS$_4$ factor have been studied extensively using Lax pairs in \cite{Bena:2003wd,Arutyunov:2008if,Stefanski:2008ik,Sorokin:2010wn,Zarembo:2010sg,Frolov:2005dj}. 
On the other hand, a popular approach towards disproving classical integrability is based on the Kovacic's algorithm \cite{Roychowdhury:2017vdo,Nunez:2018ags,Nunez:2018qcj,K1,B1,K2} that has recently found extensive applications in the context of gauge/string duality.

Together with the above, it is also equally important to examine the integrable structure of the ``deformed" superconformal field theories using a holographic setup.
Among these theories, $\beta$-deformed $\mathcal N =4$ super Yang-Mills (SYM) \cite{Lunin:2005jy} and $\gamma$-deformed Klebanov-Witten model \cite{Panigrahi:2016zny} are worthwhile to be mentioned.
It is shown that the real $\beta$-deformed $\mathcal N =4$ SYM gives rise to an integrable structure \cite{Frolov:2005ty,Frolov:2005dj}, while it looses integrability for imaginary $\beta$ deformation \cite{Giataganas:2013dha}.
On the other hand, the deformed Klebanov-Witten exhibits a chaotic behaviour whose phase space dynamics is controlled by some deformation parameter supplemented by the energy of the system \cite{Panigrahi:2016zny}.

Following the above spirit, in the present paper, we are interested in studying the integrability and/or non-integrability of \emph{marginally} deformed 4d $ \mathcal{N}=2 $ SCFTs that are introduced recently in \cite{Nunez:2019gbg}. 
The marginal deformed quiver corresponding to the dual $\gamma$-deformed background is conjectured to be a superconformal fixed point that preserves $ \mathcal{N} =1$ supersymmetries. 
It is shown that the dual $ \gamma $-deformed supergravity background can be obtained by applying an $ SL(3,R)$ transformation in the eleven dimensional M-theory solution followed by a type-IIA reduction \cite{Lunin:2005jy,Nunez:2019gbg,Gursoy:2005cn,Gauntlett:2005jb}. 
In what follows, we explore semi-classical string solutions associated to these class of $\gamma$-deformed supergravity backgrounds and estimate various chaos indicators.
Based on our findings and the holographic principle, this would eventually allow us to conjecture about the integrability and/or non-integrability of the dual $ \mathcal{N} =1$ quiver at strong coupling.

The equations of motion of semi-classical strings are in principle difficult to solve analytically.
Therefore, we solve them numerically, upon fixing the winding numbers associated with the remaining  $U(1)$ isometries. 
We analyse the Poincar\'{e} sections and estimate the corresponding Lyapunov exponents for different values of $\gamma$ and show that the chaotic dynamics of the string is explicitly controlled by the deformation parameter $\gamma$.

Upon fixing the winding numbers $\lambda = k = 1$, we study various Poincar\'{e} sections and the associated Lyapunov exponents\footnote{One can show that for winding numbers greater than unity, the plots do not change qualitatively.}.
We notice that for large values of the deformation parameter ($\gamma >>1$), the system eventually becomes more and more chaotic.
On the other hand, the string does not exhibit any chaotic behaviour when the deformation is small enough ($\gamma \sim 0$).  
This clearly connects to the fact that the integrable structure persists for the usual Sfetsos-Thompson (ST) background \cite{Sfetsos:2010uq,Itsios:2013wd,Nunez:2018qcj} which is a special case of $\gamma = 0$. 
From the perspective of the bulk theory, the above observation allows us to conjecture that the $\gamma$-deformation acts as an interpolation between an integrable $\mathcal N = 2$ SCFTs ($\gamma =0$) and non-integrable $\mathcal N=1$ SCFTs ($\gamma >>1$) in the dual QFT description\footnote{Notice that, unlike the ST example \cite{Nunez:2018qcj}, here we have a much richer structure in the sense that for small and non-zero $\gamma$, we have a class of type-IIA backgrounds that preserves integrability.}.

\vskip .1in

The rest of the paper is organised as follows. 
We briefly 
review the $ \gamma $-deformed type-IIA background in section (\ref{sec2}). 
In section (\ref{sec3}), we study the Poincar\'{e} sections and the Lyapunov exponents taking specific examples of $\gamma$-deformed type-IIA solutions. 
Finally, we draw our conclusion in section (\ref{sec4}), where we provide physical explanation of our observations.

%%%%%%%%%%%%%%%%%%%%%%%%%%%%%%%%%%%%%%%%%%%%%%%%%%%
\section{Marginal deformations of 4d $\mathcal N = 2$ SCFTs  }  \label{sec2}

Here, we first review the marginal deformations of four dimensional $ \mathcal{N}=2 $ SCFTs \cite{Nunez:2019gbg}.
The marginal deformation of these theories results in a new class of $\mathcal{N}=1$ SCFTs in four spacetime dimensions.
The holographic dual of which corresponds to a $\gamma$-deformed type-IIA supergravity which was constructed following the prescriptions of \cite{Lunin:2005jy,Gursoy:2005cn,Gauntlett:2005jb,Gaiotto:2009we}.

These new class of $\gamma$-deformed backgrounds are obtained following an $SL(3,R)$ transformation in eleven dimensional M-theory background while keeping $\gamma$ as a deformation parameter.
Upon dimensional reduction along one of the $U(1)$ isometric directions \footnote{See \cite{Nunez:2019gbg} for details.}, one finds a ten dimensional type-IIA background of the form \cite{Nunez:2019gbg}
\begin{eqnarray} \label{dgm}
&ds^2_{{\text{IIA}}} =&   \alpha^\prime \mu^2 \Bigg[ 4 f_1 ds_{AdS_5}^2   + f_2  \Big(d\sigma^2 + d\eta^2\Big) + f_3 d\chi^2  + \frac{f_3 \sin^2\chi}{1+ \gamma^2 f_3 f_4 \sin^2 \chi} d\xi^2  \cr
&& +  \ \frac{f_4 }{1+ \gamma^2 f_3 f_4  \sin^2 \chi} \Big(d\beta - \gamma f_5 \sin\chi d\chi\Big)^2\Bigg] \ . 
 \end{eqnarray}
In the limit  $ \gamma  \rightarrow 0 $, the above ten dimensional solution \eqref{dgm} maps into the standard $\mathcal{N}=2$ supersymmetric Gaiotto-Maldacena background \cite{Reid-Edwards:2010vpm,Aharony:2012tz,Gaiotto:2009gz,Lozano:2016kum,Nunez:2019gbg,Nunez:2018qcj}. 

In global coordinates, the $ AdS_5 $ line element can be expressed as
\begin{eqnarray} \label{ads5}
ds_{AdS_5}^2 =  -\cosh^2 r dt^2 + dr^2 + \sinh^2r d\Omega_3^2 \ ,
 \end{eqnarray}
 where $d\Omega_3^2$ is the metric of the three-sphere with unit radius. 

The warp factors
 $ f_i (\sigma , \eta) $s in \eqref{dgm} can be expressed in terms of a potential function $V(\sigma, \eta)$  \cite{Nunez:2019gbg}
\begin{eqnarray} \label{f}
&&f_1 = \Bigg(\frac{2\dot{V}-\ddot{V}}{ V^{\prime \prime}}\Bigg)^{\frac{1}{2}}~;~ f_2 = f_1 \frac{2V^{\prime\prime}}{\dot{V}}~;~f_3 =  f_1 \frac{2V^{\prime\prime}\dot{V}}{\Delta}~;~ f_4 = f_1 \frac{4V^{\prime\prime}}{2\dot{V}-\ddot{V}} \sigma^2 \cr
&&f_5 = 2 \Big(\frac{\dot{V}\dot{V}^\prime}{\Delta} - \eta\Big)~;~f_6 = \frac{2\dot{V}\dot{V}^\prime}{2\dot{V}-\ddot{V}}~;~f_7 =  - \frac{4 \dot{V}^2 V^{\prime\prime}}{\Delta}~;~f_8 = \Bigg[ \frac{4 \big(2\dot{V}-\ddot{V}\big)^3}{ \mu^{12} V^{\prime\prime}\dot{V}^2 \Delta^2}\Bigg]^{\frac{1}{2}} \ , 
 \end{eqnarray}
which satisfies Laplace's equation of the form 
\begin{eqnarray} 
\ddot V + \sigma^2  V^{\prime\prime} =  0 \ . 
 \end{eqnarray}
In the expressions of  $f_i$ and $\Delta$, the dot and the prime of the potential function $ V(\sigma , \eta) $ can be explicitly written as 
\begin{eqnarray} 
&& \dot{V} = \sigma \partial_\sigma V \ ,  \  V^\prime = \partial_\eta V \ ,  \  \ddot{V} =  \sigma \partial_\sigma \dot{V} \ ,    \  V^{\prime\prime} = \partial_\eta^2 V \ ,   \  \dot{V}^\prime =  \sigma \partial_\sigma V^\prime \ ,  \cr
&&\Delta =  \big(2\dot{V} - \ddot{V} \big) V^{\prime\prime} + \big(\dot{V}^\prime\big)^2 . 
\end{eqnarray}

The $\gamma$-deformed type-IIA background in \eqref{dgm} is supported by NS-NS two form and RR one form and three form fluxes along with the background dilaton \footnote{For the purpose of present analysis, we restrict ourselves only to the NS sector of the full superstring background. We also set $\alpha^\prime = \mu = 1$ in our calculation.}  \cite{Nunez:2019gbg}
\begin{eqnarray} \label{fie}
B_2 &=& \frac{\mu^2 \alpha^\prime }{1+ \gamma^2 f_3 f_4  \sin^2 \chi}  \bigg(f_5 d\Omega_2 (\chi, \xi) - \gamma f_3 f_4  \sin^2\chi d\xi \wedge d\beta\bigg) \ ,\cr
 C_1 &=& \mu^4 \sqrt{\alpha^\prime}  \  \bigg(f_6 d\beta + \gamma \big(f_7 - f_5 f_6\big) \sin\chi d\chi\bigg) \ ,\cr
 C_ 3 &=&   \frac{\mu^6 \alpha^{\prime \frac{3}{2}}}{1+ \gamma^2 f_3 f_4 \sin^2 \chi} f_7 d\beta \wedge d\Omega_2 (\chi, \xi) \ ,\cr
e^{2 \Phi} &=& \frac{ f_8}{  1+ \gamma^2 f_3 f_4 \sin^2 \chi} \ . 
\label{dgmf}
 \end{eqnarray}

\section{String motion in marginal deformed background   }  \label{sec3}

\subsection{Sigma model}

We begin our analysis with bosonic strings propagating on a curved manifold endowed with a metric $G_{\mu \nu}$ together with  NS-NS two-form field $B_{\mu \nu}$.
The dynamics of the string is characterised by the {\it{Polyakov}} action
\begin{eqnarray} \label{action}
S= - \frac{1}{2} \int d\tau d \tilde \sigma  \bigg[\eta^{\alpha \beta} G_{\mu \nu}  + \epsilon^{\alpha \beta} B_{\mu \nu} \Bigg] \partial _\alpha X^\mu  \partial _\beta X^\nu \ , 
 \end{eqnarray}
where $\eta_{\alpha \beta}$ is the two dimensional world sheet metric of the form $- \eta_{\tau \tau} =  \eta_{\tilde \sigma \tilde \sigma} = 1$ and $X^\mu$ are the string embedding coordinates on the world sheet. 
Moreover, we set our convention as $- \epsilon_{\tau \tilde \sigma} =  \epsilon_{\tilde \sigma \tau} = - 1$, throughout the rest of the analysis.

The canonical momentum corresponding to the coordinate $X^\mu$ takes the form
\begin{eqnarray} \label{mom}
p_\mu = G_{\mu \nu} \partial_\tau X^\nu + B_{\mu \nu} \partial_{\tilde \sigma} X^\nu \ . 
 \end{eqnarray}

The Hamiltonian of the system is given by
\begin{eqnarray} \label{ham}
\mathcal H = p_\mu \dot X^\mu - \mathcal L =  \frac{1}{2} G_{\mu \nu} \Big(\partial_\tau X^\mu \partial_\tau X^\nu + \partial_{\tilde \sigma}  X^\mu \partial_{\tilde \sigma}  X^\nu\Big) \ . 
\end{eqnarray}

The equations of motion for $X^\mu$ is supplemented by the Virasoro constraints that leads to the vanishing of the 2d stress tensor 
\begin{eqnarray} \label{vira}
T_{\tau \tau} = \frac{1}{2} G_{\mu \nu} \Big(\partial_\tau X^\mu \partial_\tau X^\nu + \partial_{\tilde \sigma} X^\mu \partial_{\tilde \sigma} X^\nu\Big)  = 0  \  ;  \  T_{\tau \tilde\sigma} =  \frac{1}{2} G_{\mu \nu} \partial_\tau X^\mu \partial_{\tilde \sigma}  X^\nu = 0  \ . 
 \end{eqnarray}
Notice that, from \eqref{vira}, it is straightforward to show that the Hamiltonian \eqref{ham} could be read as the time-time component of the 2d stress tensor namely $\mathcal H = T_{\tau \tau}$.

To begin with, we consider that the string sits at the center ($r=0$) of the AdS$_5$ and wraps the $U(1)$ isometries of the $\gamma$-deformed background. 
We choose an ansatz of the following form 
\begin{eqnarray} \label{embd}
&&t = t(\tau)  \ ;  \  \sigma = \sigma (\tau) \ ; \ \eta = \eta (\tau) \ ;  \  \chi = \chi (\tau) \ , \cr
 && \xi = \xi(\tilde \sigma) = k \tilde \sigma  \ ; \  \beta = \beta(\tilde \sigma) = \lambda \tilde \sigma \ . 
  \end{eqnarray}
Here $\{\tau, \tilde\sigma\}$ are the coordinates of the two-dimensional world sheet and $\{k, \lambda\}$ are the integers which we denote as the winding numbers of the string along the $U(1)$ isometric directions ($\xi$ and $\beta$) of the $\gamma$-deformed background \eqref{dgm}. 
Following the embedding \eqref{embd},
 the world sheet Lagrangian takes the following form 
\begin{eqnarray} \label{ws-1}
\mathcal L_{P} &=& - \frac{1}{2} \Bigg[4f_1 \dot t^2 - f_2 \Big(\dot \sigma^2 + \dot \eta^2 \Big) - \Big(f_3 + \gamma^2 \frac{f_4 f_5^2 }{1+ \gamma^2 f_3 f_4  \sin^2 \chi}  \sin^2\chi \Big)  \dot \chi^2 +  \frac{f_3 k^2 \sin^2\chi}{1+ \gamma^2 f_3 f_4 \sin^2 \chi} \cr
&&+ \ \frac{f_4 \lambda^2}{1+ \gamma^2 f_3 f_4  \sin^2 \chi} \Bigg]+  \frac{k f_5 }{1+ \gamma^2 f_3 f_4  \sin^2 \chi}  \sin\chi  \ \dot\chi \ , 
  \end{eqnarray}
where dot denotes the derivative with respect to the world sheet time ($\tau$).

The Lagrangian \eqref{ws-1}, leads to the following Hamiltonian 
\begin{eqnarray} \label{Ham:Gen}
 \mathcal H &=& \frac{1}{2} \Bigg[-\frac{p_t^2}{4f_1} + \frac{1}{f_2} \Big(p_\sigma^2 + p_\eta^2\Big) 
 + \frac{f_3 k^2 \sin^2\chi}{1+ \gamma^2 f_3 f_4 \sin^2 \chi} + \frac{f_4 \lambda^2}{1+ \gamma^2 f_3 f_4  \sin^2 \chi} \cr
&&+  \Big(f_3 + \gamma^2 \frac{f_4 f_5^2 }{1+ \gamma^2 f_3 f_4  \sin^2 \chi}  \sin^2\chi \Big)^{-1}  \Big(p_\chi  -  \frac{k f_5 }{1+ \gamma^2 f_3 f_4  \sin^2 \chi}  \sin\chi \Big)^2 \Bigg]  \ . 
\end{eqnarray}

\subsection{ Example I : $\gamma$-deformed Abelian T-dual  }

The potential function for the Abelian T-dual (ATD) case takes the form \cite{Lozano:2016kum}
\begin{eqnarray} \label{V atd}
V_{\text{ATD}} (\sigma,\eta) = \ln \sigma - \frac{1}{2} \sigma^2 + \eta^2 \ . 
 \end{eqnarray}
Using \eqref{V atd}, the associated functions $f_i(\sigma,\eta)$ in \eqref{f} take the following form 
\begin{eqnarray} \label{f:atd}
&&f_1 =1 \ ; \ f_2 = \frac{4}{1-\sigma^2} \ ; \ f_3 =  1-\sigma^2 \ ;  \ f_4 =4 \sigma^2 \ ,  \cr
&&f_5 = - 2\eta \ ; \ f_6 = 0 \ ; \ f_7 =  - 2 \big(1-\sigma^2\big)^2 \ ;  \ f_8 = \frac{64}{1-\sigma^2} \ . 
 \end{eqnarray}
Using \eqref{f:atd}, the Hamiltonian \eqref{Ham:Gen} can be expressed as
%%%%
\begin{align}\label{Ham:atd}
\begin{split}
\mathcal{H} &= \frac{1}{2}\Bigg[-\frac{p_{t}^{2}}{4}
+\frac{\qty(1-\sigma^{2})}{4}\qty(p_{\sigma}^{2}+p_{\eta}^{2})
+\frac{\qty(1-\sigma^{2})k^{2}\sin^{2}\chi + 4\sigma^{2}
\lambda^{2}}{1+4\gamma^{2}\sigma^{2}\qty(1-\sigma^{2})\sin^{2}
\chi}
\\[6pt]
& + \qty(p_{\chi}+\frac{2k\eta \sin\chi}{1+4\gamma^{2}
\sigma^{2}\qty(1-\sigma^{2})\sin^{2}\chi})^{2} \qty(\qty(1-
\sigma^{2})+\frac{16 \eta^{2}\sigma^{2}\gamma^{2}\sin^{2}\chi}{1
+4\gamma^{2}\sigma^{2}\qty(1-\sigma^{2})\sin^{2}\chi})^{-1} \Bigg]  \, ,
\end{split}
\end{align}
%%%%
which serves as the basis for studying the Hamiltonian dynamics and hence the various chaos indicators that we compute next. 

Using \eqref{Ham:atd}, the Hamilton's equations of motion could be read as 
%%%%
\begin{subequations}\label{EoM:HAb}
\begin{alignat}{2}
\begin{split}
\dot{\chi} &=\frac{p_{\chi}}{1-\sigma^{2}}  \, ,
\end{split}
\label{HAb:chi}\\[5pt]
\begin{split}
\dot{p}_{\chi} &= \frac{\qty(\sigma^{2}-1)\qty(k^{2}-16\gamma^{2}
\lambda^{2}\sigma^{4})}{2\qty(1+4\gamma^{2}\sigma^{2}\qty(1-
\sigma^{2})\sin^{2}\chi)^{2}}\sin 2\chi   \, ,
\end{split}
\label{HAb:pchi}\\[5pt]
\begin{split}
\dot{\sigma} &= \frac{1}{4}\qty(1-\sigma^{2})p_{\sigma} \, ,
\end{split}
\label{HAb:sig}\\[5pt]
\begin{split}
\dot{p}_{\sigma} &=  \frac{1}{4}\sigma \Bigg[ p^{2}_{\sigma}
-\frac{4p^{2}_{\chi}}{\qty(1-\sigma^{2})^{2}}+\frac{4
\qty(k^{2}\sin^{2}\chi -4 \lambda^{2})}{1+4\gamma^{2}
\sigma^{2}\qty(1-\sigma^{2})\sin^{2}\chi}   
\\[4pt]
& \ +\frac{16\gamma^{2}\qty(2\sigma^{2}-1)\sin^{2}\chi
\qty(k^{2}\qty(\sigma^{2}-1)\sin^{2}\chi-4\lambda^{2}
\sigma^{2})}{\qty(1+4\gamma^{2}\sigma^{2}\qty(1-\sigma^{2})
\sin^{2}\chi)^{2}}\Bigg]  \, ,
\end{split}  \label{HAb:psig}
\end{alignat}
\end{subequations}
%%%%

while in writing the above set equations of motion \eqref{HAb:chi}-\eqref{HAb:psig}, we choose $\eta =0$ and $p_{\eta}=0$ for our convenience.

\begin{figure}[h]\label{fig:AbPS}
     \centering
     \begin{subfigure}[b]{0.4\textwidth}
         \centering
         \includegraphics[width=\textwidth]{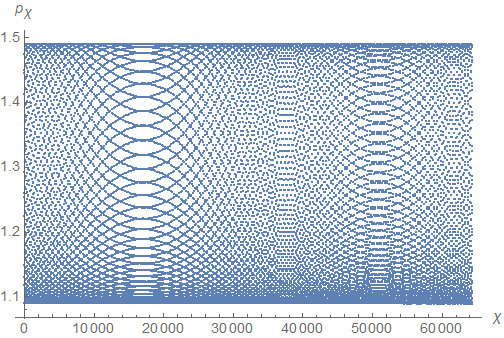}
         \caption{$E=3$, $\gamma=0.25$. Initial data: $\sigma=0.1$, $\chi=0.5$,
         $p_{\sigma}=0$, $p_{\chi}=1.40099$.}
         \label{fig:AbPS1}
     \end{subfigure}
     \hfill
     \begin{subfigure}[b]{0.4\textwidth}
         \centering
         \includegraphics[width=\textwidth]{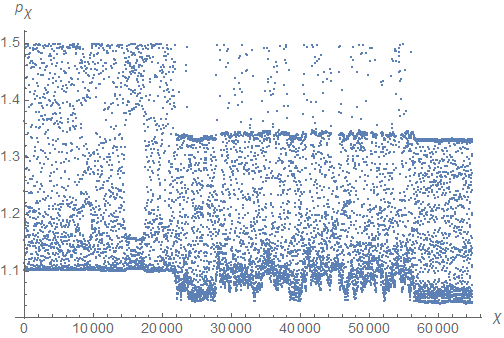}
         \caption{$E=3$, $\gamma=2$. Initial data: $\sigma=0.1$, $\chi=0.5$,
         $p_{\sigma}=0$, $p_{\chi}=1.40425$.}
         \label{fig:AbPS2}
     \end{subfigure}
     \hfill
     \begin{subfigure}[b]{0.4\textwidth}
         \centering
         \includegraphics[width=\textwidth]{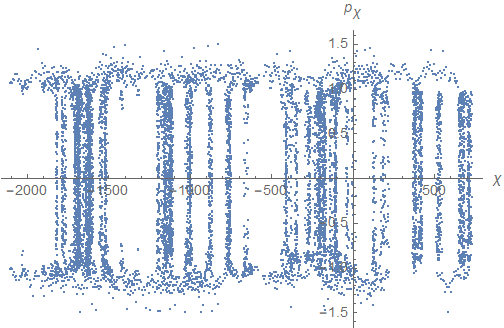}
         \caption{$E=3$, $\gamma=3$. Initial data: $\sigma=0.1$, $\chi=0.5$,
         $p_{\sigma}=0$, $p_{\chi}=1.40808$.}
         \label{fig:AbPS3}
     \end{subfigure}
     \hfill
     \begin{subfigure}[b]{0.4\textwidth}
     \centering
     \includegraphics[width=\textwidth]{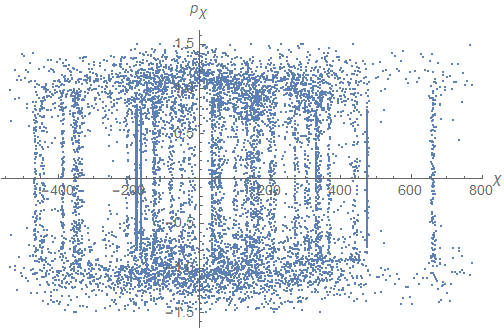}
        \caption{$E=3$, $\gamma=5$. Initial data: $\sigma=0.1$, $\chi=0.5$,
         $p_{\sigma}=0$, $p_{\chi}=1.41835$.}
        \label{fig:AbPS4}
      \end{subfigure}  
        \caption{Plots of Poincar\'{e} sections for $\gamma$-deformed
        Abelian T-dual background. Here we fix the energy of the string
       $E=3$ and choose different values of the deformation parameter
       $\gamma$. As the deformation increases, the configuration becomes
       more and chaotic.}
\end{figure}

We now numerically estimate the Poincar\'{e} sections by solving the Hamilton's equations of motion \eqref{HAb:chi}-\eqref{HAb:psig} that is subjected to the Virasoro constraints \eqref{vira}. 
In our numerical plots, we set the energy of the string to be $E=3$. On the other hand, we choose the following set of values for the deformation parameter: $\{\gamma\}=\{0.25, 2.0,3.0,5.0\}$. The corresponding plots are found in Figs.(\ref{fig:AbPS1})-(\ref{fig:AbPS4}). Also notice that, in our analysis we set the winding numbers as $k=1$ and $\lambda =1$ (for general values of the winding numbers see Appendix (\ref{ap A})).

We now proceed to choose sets of initial conditions that generate solutions to the dynamical equations \eqref{HAb:chi}-\eqref{HAb:psig} for different values of the deformation parameter $\gamma$. These initial conditions are indeed consistent with the Virasoro constraints \eqref{vira}. In the following analysis, we choose the initial conditions as $\sigma(0)=0.1$ and $p_{\sigma}(0)=0$. With this initial set of data and the choice $\chi(0)=0.5$, the corresponding $p_{\chi}(0)$ is fixed while satisfying the constraint \eqref{vira}. For different values of the deformation parameter $(\gamma)$ this is shown in table \ref{Tab:AbT} below. 
%%%%
\begin{table}[h!]
\centering
 \begin{tabular}{||c c c c c||} 
 \hline
 $\gamma$ & $\chi$ & $\sigma$ & $p_{\sigma}$  & $p_{\chi}$ \\ [0.5ex] 
 \hline\hline
 0.25 & 0.5 & 0.1 & 0 & 1.40099 \\ 
 2.0 & 0.5 & 0.1 & 0 & 1.40425 \\
 3.0 & 0.5 & 0.1 & 0 & 1.40808 \\
 5.0 & 0.5 & 0.1 & 0 & 1.41835 \\ [1ex] 
 \hline
 \end{tabular}
 \caption{Initial conditions for different values
 of the deformation parameter $\gamma$ for the Abelian
 T-dual case.}
 \label{Tab:AbT}
\end{table}
%%%%

In the present example, the phase space under consideration is four dimensional and characterised by the following set of generalised coordinates and momenta: $\{\sigma,\chi,p_{\sigma},p_{\chi} \}$. At fixed energy $E=3$ of the string, for small deformation ($\gamma=0.25$) the phase phase trajectory is observed to be quasi-periodic (cf. Fig.(\ref{fig:AbPS1})). However, as the deformation is increased the perturbations of the Hamiltonian result the destruction of the quasi-periodicity and randomly distributed data sets are generated. These are shown in Figs.(\ref{fig:AbPS2})-(\ref{fig:AbPS4}). Since these distributions lack any nice foliation in the form of KAM tori \cite{PandoZayas:2010xpn,Basu:2011dg,Basu:2011di}, which are observed in an integrable systems, we conclude that the system is chaotic at large deformations.  

%%%%
%%%%

%%%%
\begin{figure}[h!]
\centering
	\includegraphics[width=0.6\textwidth]{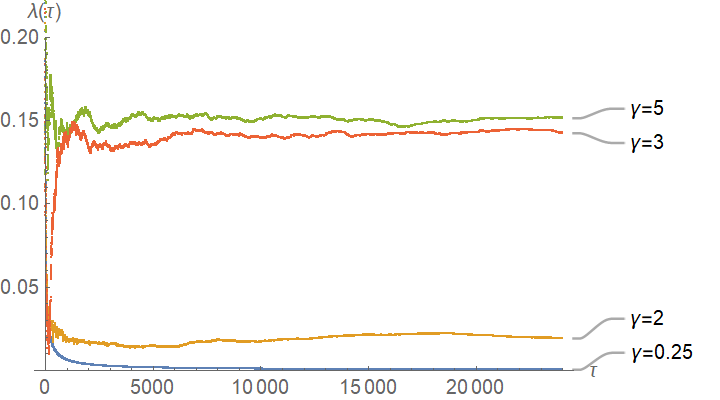}
        \caption{Numerical plots of Lyapunov exponent(s) for
        $\gamma$-deformed Abelian T-dual background. Here
        we fix the energy of the string as $E=3$. Different curves
        correspond to different values of the deformation parameter
        $\gamma$, as indicated in the figure.}
     \label{fig:AbLE}
\end{figure}
%%%%

Next we numerically compute another chaos indicator namely, the Lyapunov exponent \cite{PandoZayas:2010xpn,Basu:2011dg,Basu:2011di}. During the time evolution of a chaotic system, the system becomes sensitive to the choice of initial conditions that are imposed on the dynamics. The Lyapunov exponent ($\lambda$) measures the deviation between two nearby trajectories in the phase space due to a small variation in the initial conditions.\footnote{The Lyapunov exponent is defined as  \cite{PandoZayas:2010xpn,Basu:2011dg,Basu:2011di}
%%%%
\begin{equation}\label{Def:Lyap}
\lambda = \lim_{\tau\rightarrow\infty}\lim_{\Delta X_{0}\rightarrow\infty}
\frac{1}{\tau}\log\frac{\Delta X\qty(X_{0},\tau)}{\Delta X\qty(X_{0},0)} \, ,
\end{equation}
%%%%
where $\Delta X\qty(X_{0},\tau)$ is the separation between the infinitesimally close trajectories in the phase space at sufficiently late times.}

As before, we choose the energy of the string as $E=3$, whereas the rest of the initial conditions are chosen as in table (\ref{Tab:AbT}).
With this initial set of data, we study the dynamical evolution of two nearby orbits in the phase space which have an initial separation $\Delta X (0) = 10^{-7}$ (cf. \eqref{Def:Lyap}). The resulting plots are shown in Fig.(\ref{fig:AbLE}). We observe that, for $\gamma >1$, the Lyapunov exponent saturates to positive non-zero values at sufficient late times indicating the onset of chaos.
On the other hand, for small $\gamma <1$, it asymptotes to zero at sufficiently late times indicating an underlying non-chaotic dynamics of the associated phase space.

\subsection{Example II : $\gamma$-deformed Sfetsos-Thompson solution }\label{Sec:NAbT}

For Sfetsos-Thompson (ST) solution (also known as non-Abelian T-dual (NATD) solution)  the corresponding potential function reads as \cite{Lozano:2016kum}
%%%%
\begin{equation} \label{V:natd}
V_{\text{ST}} (\sigma,\eta) = \eta\Big(\ln \sigma - \frac{1}{2}
\sigma^2\Big) + \frac{1}{3} \eta^3 \ . 
\end{equation}
%%%%
Using \eqref{V:natd}, the associated metric functions $f_i(\sigma,\eta)$ in \eqref{f} take the following form
%%%% 
\begin{eqnarray} \label{f:natd}
&&f_1 =1 \ ; \ f_2 = \frac{4}{1-\sigma^2} \ , \ f_3 =  \frac{4\eta^2
\big(1-\sigma^2\big)}{4\eta^2 + \big(1-\sigma^2\big)^2} \ ,  \ f_4 =4 \sigma^2 \ ,
\cr
&&f_5 = - \frac{8\eta^3}{4\eta^2 + \big(1-\sigma^2\big)^2} \ , \ f_6 =
\big(1-\sigma^2\big)^2 \ , \ f_7 =   - \frac{8\eta^3\big(1-\sigma^2\big)^2}{4\eta^2 + \big(1-\sigma^2\big)^2} \ ,
\cr
&&f_8 =  \frac{256}{\big(1-\sigma^2\big)^2 \big(4\eta^2 + \big(1-\sigma^2\big)^2\big)} \ . 
\end{eqnarray}
%%%%

Using \eqref{f:natd} the Hamiltonian of the system \eqref{Ham:Gen} can be expressed as  \footnote {Notice that, in writing the above equations of motion \eqref{HNab:sig}-\eqref{HNab:pchi} we set $\eta=1$ and $p_{\eta}=0$.}

%%%%
\begin{align}\label{Ham:natd}
	\mathcal{H}&=\frac{1}{8} \Bigg[-p_t^2-\big(\sigma ^2-1\big)
	p_\sigma^2-\big(\sigma ^2-1\big)
	p_\eta^2\nonumber\\&-\frac{16 \eta ^2 \big(\sigma
		^2-1\big) \left(\frac{k^2 \sin ^2\chi
		}{1-\frac{16 \gamma ^2 \eta ^2 \sigma ^2
				\big(\sigma ^2-1\big) \sin ^2\chi }{4 \eta
				^2+\big(\sigma ^2-1\big)^2}}+\frac{\big(8
			\eta ^3 k \sin \chi +4 \eta ^2 p_\chi
			\big(-4 \gamma ^2 \sigma ^4 \sin ^2\chi +4
			\gamma ^2 \sigma ^2 \sin ^2\chi
			+1\big)+\big(\sigma ^2-1\big)^2 p_\chi\big)^2}{16 \eta ^4 \big(\sigma ^2
			\big(16 \gamma ^2 \eta ^2 \sin ^2\chi
			-1\big)+1\big)^2}\right)}{4 \eta
		^2+\big(\sigma ^2-1\big)^2}\nonumber\\&+\frac{16 \sigma
		^2 \left(\lambda ^2+\frac{4 \gamma ^2 \eta ^2
			\sin ^2\chi  \big(8 \eta ^3 k \sin \chi +4
			\eta ^2 p_\chi \big(-4 \gamma ^2 \sigma ^4
			\sin ^2\chi +4 \gamma ^2 \sigma ^2 \sin
			^2\chi +1\big)+\big(\sigma ^2-1\big)^2
			p_\chi\big)^2}{\big(4 \eta
			^2+\big(\sigma ^2-1\big)^2\big)^2
			\big(\sigma ^2 \big(16 \gamma ^2 \eta ^2 \sin
			^2\chi
			-1\big)+1\big)^2}\right)}{1-\frac{16 \gamma
			^2 \eta ^2 \sigma ^2 \big(\sigma ^2-1\big)
			\sin ^2\chi}{4 \eta ^2+\big(\sigma
			^2-1\big)^2}}\Bigg] \, .
\end{align}
%%%%

The resulting Hamilton's equations of motion easily follow from \eqref{Ham:natd} and can be summarised as
%%%%
\begin{subequations}
\begin{alignat}{2}
\begin{split}
	\dot{\sigma}&=-\frac{1}{4}(-1+\sigma^2)p_\sigma \, ,
\end{split}\label{HNab:sig}
\\[6pt]
\begin{split}
	\dot{p}_\sigma &=\frac{1}{4\sigma^3}\bigg(\sigma ^4
	\big(p_\sigma^2+p_\chi^2\big)-\frac{16 \lambda ^2
	\big(\sigma ^2-3\big)
		\big(\sigma ^2+1\big) \sigma ^6}{\big(\sigma
		^2-1\big) \big(\sigma ^2 \big(8 \gamma ^2
		\big(\sigma ^2-1\big) \cos 2 \chi -8 \gamma
		^2 \sigma ^2+8 \gamma ^2+\sigma
		^2-2\big)+5\big)}
		\\
		&+\frac{16 \lambda ^2
		\big(\sigma ^4-10 \sigma ^2+5\big)
		\big(\sigma ^4-2 \sigma ^2+5\big) \sigma
		^4}{\big(\sigma ^2-1\big) \big(\sigma ^2
		\big(8 \gamma ^2 \big(\sigma ^2-1\big) \cos
		2 \chi -8 \gamma ^2 \sigma ^2+8 \gamma
		^2+\sigma
		^2-2\big)+5\big)^2}+\frac{k^2}{\gamma
		^2}
		\\
		&+\frac{k^2-\sigma ^2 \big(k^2+4 \gamma ^2
		p_\chi \big(4 k \sin \chi +p_\chi\big)\big)}{\gamma ^2
		\big(\sigma ^2
		\big(8 \gamma ^2 \cos (2 \chi )-8 \gamma
		^2+1\big)-1\big)^2}+\frac{2 k^2-\sigma ^2
		\big(k^2+4 \gamma ^2 p_\chi \big(4 k \sin
		\chi +p_\chi\big)\big)}{\gamma ^2
		\big(\sigma ^2 \big(8 \gamma ^2 \cos (2 \chi
		)-8 \gamma ^2+1\big)-1\big)}\bigg) \, ,
\end{split}\label{HNab:psig}
\\[6pt]
\begin{split}
\dot{\chi} &=-\frac{2 k \sin \chi +p_\chi}{\sigma ^2
	\big(8 \gamma ^2 \cos 2 \chi -8 \gamma
	^2+1\big)-1}-\frac{1}{4} (\sigma
^2-1) p_\chi \, ,
\end{split}\label{HNab:chi}
\\[6pt]
\begin{split}
\dot{p}_\chi &=-\frac{2	p_\chi k \cos\chi
}{\sigma ^2 \big(8 \gamma ^2 \cos (2
	\chi )-8 \gamma ^2+1\big)-1}-\frac{32 \gamma ^2
	\lambda ^2 \big(\sigma ^6-3
	\sigma ^4+7 \sigma ^2-5\big) \sigma ^4 \sin 2
	\chi }{\big(\sigma ^2 \big(8 \gamma ^2
	\big(\sigma ^2-1\big) \cos2\chi-8 \gamma
	^2 \sigma ^2+8 \gamma ^2+\sigma
	^2-2\big)+5\big)^2}
	\\
	&+\frac{4 \cos\chi
	\big(\sigma ^2 \big(\sin\chi \big(k^2+4
	\gamma ^2 p_\chi^2\big)+k p_\chi\big)-k \big(k \sin\chi
	+p_\chi\big)\big)}{\big(\sigma ^2 \big(8 \gamma
	^2 \cos2\chi-8 \gamma
	^2+1\big)-1\big)^2} \, .
\end{split}\label{HNab:pchi}
\end{alignat}
\end{subequations}
%%%%

\begin{figure}[h!]
     \centering
     \begin{subfigure}[b]{0.4\textwidth}
         \centering
         \includegraphics[width=\textwidth]{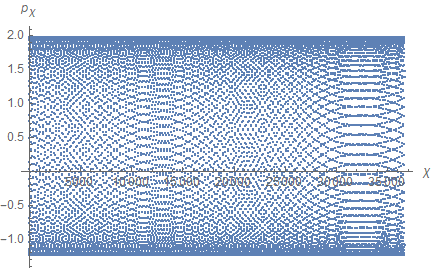}
         \caption{$E=2$, $\gamma=0.25$. Initial data: $\sigma=0.1$, $\chi=0.5$,
         $p_{\sigma}=0$, $p_{\chi}=0.0171369$.}
         \label{fig:NAbPS1}
     \end{subfigure}
     \hfill
     \begin{subfigure}[b]{0.4\textwidth}
         \centering
         \includegraphics[width=\textwidth]{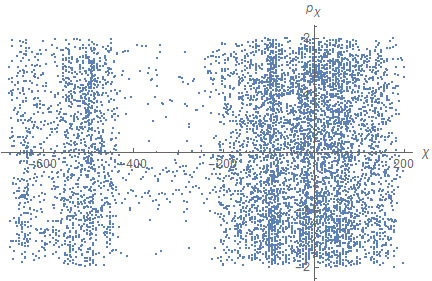}
         \caption{$E=2$, $\gamma=2$. Initial data: $\sigma=0.1$, $\chi=0.5$,
         $p_{\sigma}=0$, $p_{\chi}=0.0855865$.}
         \label{fig:NAbPS2}
     \end{subfigure}
     \hfill
     \begin{subfigure}[b]{0.4\textwidth}
         \centering
         \includegraphics[width=\textwidth]{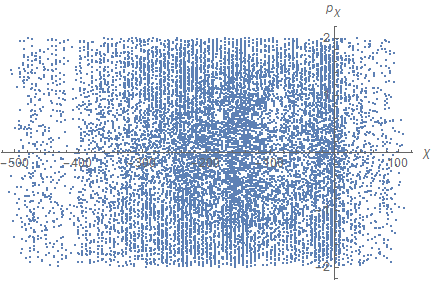}
         \caption{$E=2$, $\gamma=3$. Initial data: $\sigma=0.1$, $\chi=0.5$,
         $p_{\sigma}=0$, $p_{\chi}=0.1647749$.}
         \label{fig:NAbPS3}
     \end{subfigure}
     \hfill
     \begin{subfigure}[b]{0.4\textwidth}
     \centering
     \includegraphics[width=\textwidth]{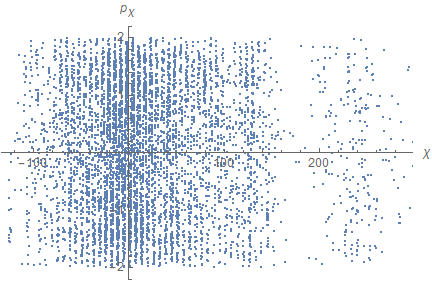}
        \caption{$E=2$, $\gamma=5$. Initial data: $\sigma=0.1$, $\chi=0.5$,
         $p_{\sigma}=0$, $p_{\chi}=0.374738$.}
        \label{fig:NAbPS4}
      \end{subfigure}  
        \label{fig:NAbPS}
        \caption{Plots of Poincar\'{e} sections for $\gamma$-deformed
        non-Abelian T-dual background. Here we fix the energy of the string
       $E=2$ and choose different values of the deformation parameter
       $\gamma$. As the deformation increases, the configuration becomes
       more chaotic.}
\end{figure}

\begin{figure}[h!]
\centering
	\includegraphics[width=0.65\textwidth]{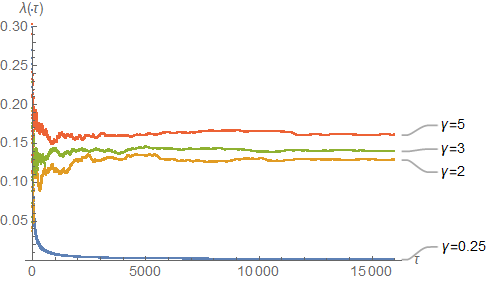}
        \caption{Numerical plots of Lyapunov exponent(s) for
        $\gamma$-deformed ST background. Here,
        we fix the energy of the string as $E=2$. Different curves
        correspond to different values of the deformation parameter
        $\gamma$, as indicated in the plot. The initial separation
        between two nearby phase space trajectories is set to be
        $\Delta X(0)=10^{-7}$.}
     \label{fig:NAbLE}
\end{figure}

In the next step, we numerically study the Poincar\'{e} sections and the Lyapunov exponents for the string configuration for different values of the deformation parameter $\gamma$. We choose the initial data as given in Table (\ref{Tab:NAbT}). These data sets are indeed consistent with the Virasoro constraints \eqref{vira}. The resulting plots are shown in Figs.(\ref{fig:NAbPS1})-(\ref{fig:NAbPS4}) and Fig.(\ref{fig:NAbLE}).

%%%%
\begin{table}[h!]
\centering
 \begin{tabular}{||c c c c c||} 
 \hline
 $\gamma$ & $\chi$ & $\sigma$ & $p_{\sigma}$  & $p_{\chi}$ \\ [0.5ex] 
 \hline\hline
 0.25 & 0.5 & 0.1 & 0 & 0.0171369 \\ 
 2.0 & 0.5 & 0.1 & 0 & 0.0855865 \\
 3.0 & 0.5 & 0.1 & 0 & 0.1647719 \\
 5.0 & 0.5 & 0.1 & 0 & 0.374748 \\ [1ex] 
 \hline
 \end{tabular}
 \caption{Initial conditions for different values
 of the deformation parameter $\gamma$ for the non-Abelian
 T-dual case. Here we set the string energy $E=2$.}
 \label{Tab:NAbT}
\end{table}
%%%%

%%%%
%%%%

%%%%
%%%%

Like before, we observe that for small values deformation parameter ($\gamma << 1$), the Poincar\'{e} section is quasi-periodic (Fig.\ref{fig:NAbPS1}) indicating an underline integrable structure. On the other hand, if the deformation is increased enough, the perturbation of the Hamiltonian destroys the quasi-periodicity and the resulting dynamics becomes chaotic. These are shown in Figs.(\ref{fig:NAbPS2})-(\ref{fig:NAbPS4}).

In Fig.(\ref{fig:NAbLE}), we plot the Lyapunov exponents ($\lambda$) for different values of the deformation parameter. 
The Lyapunov exponent for $\gamma >>1$ and at sufficiently later times, saturates to a positive value. 
On the other hand, in the limit of small deformation ($\gamma <<1$), the conclusion remains same as in the $\gamma$-deformed ATD example studied previously.
The Lyapunov asymptotes to zero indicating a non chaotic motion and this is clearly visible in the Fig.(\ref{fig:NAbLE}). 
This observation (for small range of $\gamma$) is similar to what has been observed previously in the context of ST background in \cite{Nunez:2018qcj}.

\begin{figure}[h!]
\centering
\includegraphics[width=0.7\textwidth]{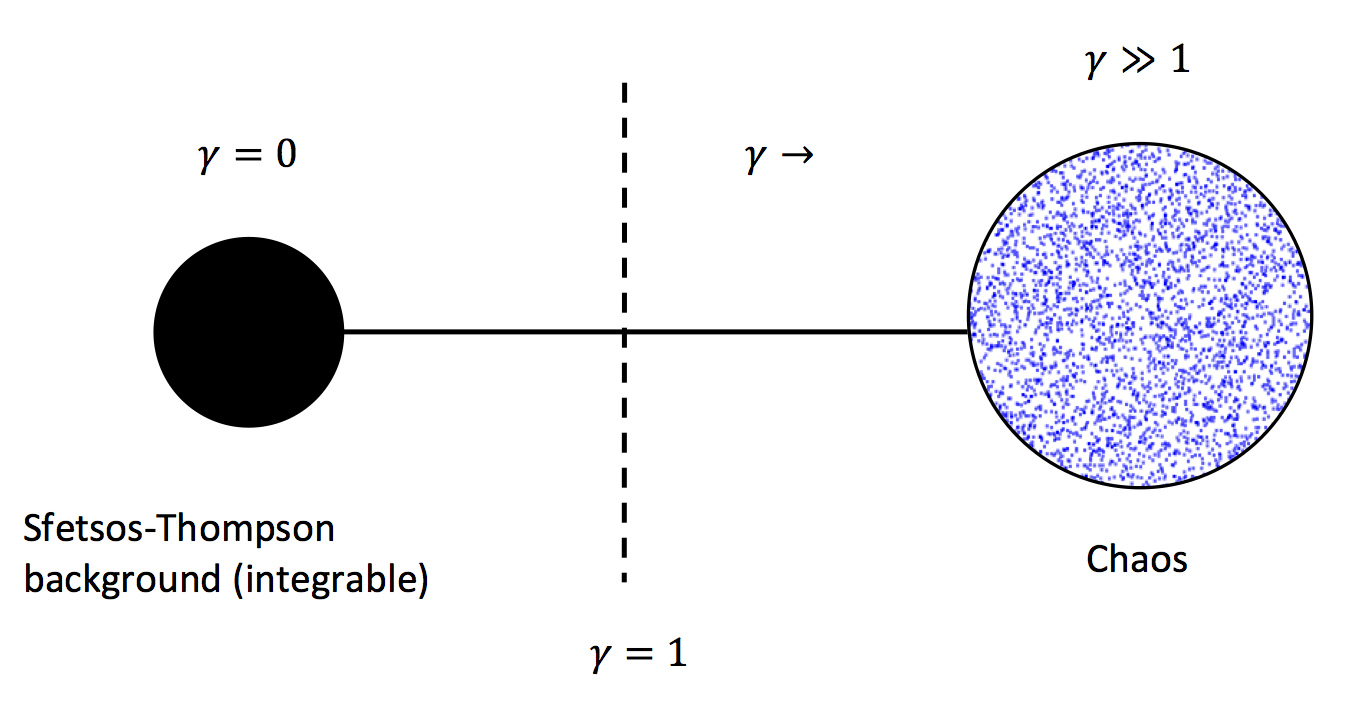}
\caption{We show $\gamma$-deformation as an interpolation
between Sfetsos-Thompson background and chaotic dynamics of strings.}
\label{fig:chaosST}
\end{figure}

Therefore, we conjecture that, for a small range of $\gamma \sim 0$, we have a class of integrable $\mathcal N =1$ backgrounds in type-IIA theory whose dual SCFTs are also conjectured to be integrable.
However, as we increase the values of $\gamma$-deformation, we deviate away from these class of integrable theories to a class of non-integrable SCFTs.
Therefore, to summarise, $\gamma$-deformation acts like a bulk interpolating parameter (see Fig.(\ref{fig:chaosST})) that connects a class of integrable $\mathcal N =1$ SCFTs (that are obtained from $\mathcal N =2$ SCFTs via marginal deformation) to a class of non-integrable $\mathcal N =1$ SCFTs at strong coupling.

\subsection{Example III : Adding flavor branes}
We now generalise our results in the presence of flavor brane \cite{Roychowdhury:2021eas}.
The first one of these backgrounds is known as the \emph{single kink} space-time where $N_{6}$ flavor $D6$-branes sit at $\eta = P$ of the internal manifold, where $P (>>1)$ is an integer. The dual SCFT endows with a gauge group $SU(N) \times SU(2N) \times \cdots \times SU(PN)$ which is closed due to an addition of $SU((P+1)N)$ flavor group. The second background that we choose is called the periodic \emph{Uluru} space-time that corresponds to placing $N_{6}$ flavor $D6$-branes at $\eta =P$ and $\eta = K+P$ of the internal manifold. These dual SCFT are characterised by a linear quiver consisting of $K$ $SU(N)$ gauge group nodes terminated on each end by placing suitable flavor nodes.

The potential function corresponding to the \emph{single kink} solution may be written as \cite{Roychowdhury:2021eas}
%%%%
\begin{equation}\label{Pot:SK}
V (\sigma \sim  0, \eta) = \eta N_6 \ln \sigma + \frac{\eta N_6
\sigma^2}{4} \Lambda_k (\eta, P) -  \frac{\eta N_6 \sigma^2}{4}
\frac{P+1}{P^2 - \eta^2} \ , 
\end{equation}
%%%%
where we define the above function as 
%%%%
\begin{align}
\begin{split} 
\Lambda_k (\eta, P) &= \big(P+1\big) \sum_{m = 1}^k
\left(\frac{1}{(2m+(2m-1)P)^2 - \eta^2} -  \frac{1}{(2m+(2m+1)P)^2
- \eta^2}\right)
\\
&\quad + \frac{P}{(2k+1)^2(1+P)^2 - \eta^2} \ . 
\end{split}
\end{align}
%%%%

\begin{figure}[h!]
	\centering
	\includegraphics[width=0.5\textwidth]{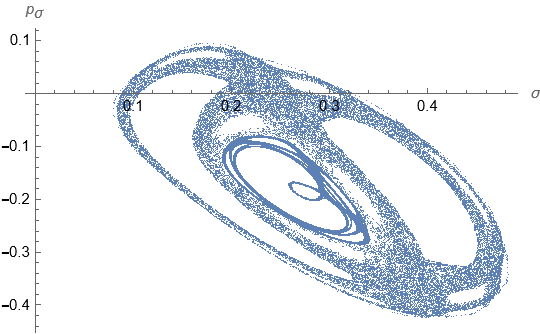}
	\caption{Plot of a Poincar\'{e} section for $\gamma$-deformed
	single kink profile. Here we fix the energy of the string
	$E=3$ and choose  of the deformation parameter
	$\gamma=0.25$.  Also, we set $\eta=7$, $p_\eta=0 $, $P=6$
	and $k=10$.}
	\label{fig:FbSk1}
\end{figure}

In what follows, we consider $\gamma$-deformation of the above theory and calculate various chaos indicators. 
In order to establish the (non-)integrability of the string sigma model, we numerically study the Poincar\'{e} sections corresponding to different values of the deformation parameter $\gamma$. The below Fig.(\ref{fig:FbSk1}) shows a Poincar\'{e} section when the value of the deformation parameter is set as $\gamma =0.25$. We set the energy of the string as $E=3$. Clearly, the Poincar\'{e} section consists of random data points in the $(\sigma-p_{\sigma})$ plane of the phase space. This indicates the chaotic dynamics of strings in the phase space.

%%%%
%%%%

Next, in the case of \emph{Uluru} space-time, the potential function may be expressed as \cite{Roychowdhury:2021eas}
%%%%
\begin{equation}\label{Pot:Ulu}
V (\sigma \sim  0, \eta) = - \eta N_6 \ln \sigma + \frac{\eta N_6 \sigma^2}{4}
\Lambda_u (\eta, K, P) +  \frac{\eta N_6 \sigma^2}{4 (P^2 - \eta^2)} \ , 
\end{equation}
%%%%
where we define 
%%%%
\begin{equation} 
\Lambda_u (\eta, K, P) = \sum_{n = 1}^u (-1)^{n+1}
\left(\frac{1}{(nK+(2n-1)P)^2 - \eta^2} -  \frac{1}{(nK+(2n+1)P)^2
-\eta^2}\right) \ . 
\end{equation}
%%%%

\begin{figure}[h!]
	\centering
	\includegraphics[width=0.55\textwidth]{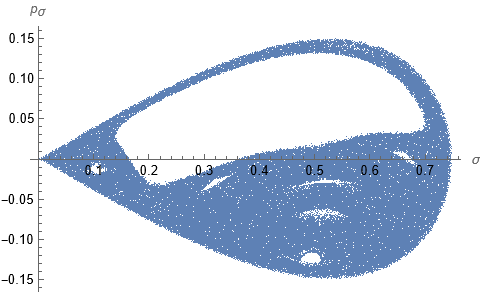}
	\caption{Plots of Poincar\'{e} sections for $\gamma$-deformed
		Uluru profile. Here we fix the energy of the string
		$E=3$ and choose  of the deformation parameter
		$\gamma=0.25$.  Also, we set $\eta=7$, $p_\eta=0 $, $P=6$,
		$n=10$ and $K=10$.}
		\label{fig:Ulu1}
\end{figure}

In this case as well we observe from fig.(\ref{fig:Ulu1}) that the Poincar\'{e} section at energy $E=3$ and deformation $\gamma=0.25$ is a random sets of data points which are indicator of chaotic dynamics of the string.
The above result is not surprising and in fact is expected since the presence of flavor branes spoil the integrability in dual SCFTs \cite{Nunez:2018qcj,Roychowdhury:2021jqt}. 

%%%%
%%%%

%%%%%%%%%%%%%%%%%%%%%%%%%%%%%%%%%%%%%%%%%%%%%%%%%%%
\section{Summary and outlook  } \label{sec4}

In this present paper, we explore chaotic dynamics for a class of $\mathcal N =1$ supergravity backgrounds namely the $\gamma$-deformed abelian and non-Abelian T-dual of Gaiotto-Maldacena solutions. 
The dual $\mathcal N =1$ SCFT is characterised by a linear quiver theory proposed in \cite{Nunez:2019gbg}. 
The deformed background can be obtained by applying an $SL(3,R)$ transformation in the eleven dimensional background followed by a type-IIA reduction. 

The primary goal was to examine the integrability of the semi-classical string trajectories in the presence of the $\gamma$-deformation as introduced in \cite{Nunez:2019gbg}. 
In our analysis, we use the standard Hamiltonian formulation and study the Poincar\'{e} sections and the Lyapunov exponents for different values of the deformation parameter ($\gamma$).

\begin{figure}[ H]
\centering
\hfill
  \begin{subfigure}{0.4\textwidth}
    \includegraphics[width=\textwidth]{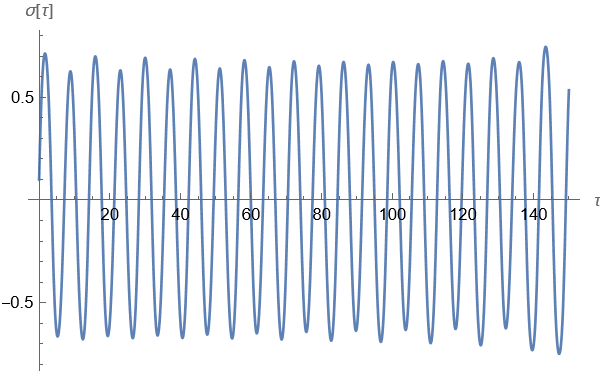}
    \caption{$\gamma=0.25$}
    \label{fig:1}
  \end{subfigure}
 \hfill
  \begin{subfigure}{0.4\textwidth}
  \centering
    \includegraphics[width=\textwidth]{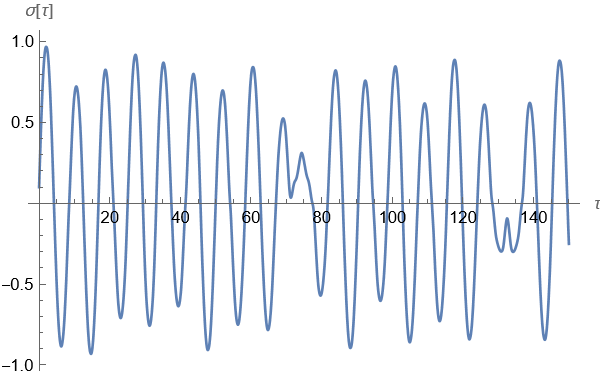}
    \caption{$\gamma=5$}
    \label{fig:2}
  \end{subfigure}
   \caption{Motion in ($\sigma-\tau$) plane for Abelian T-dual case with $E=3 \ ,  \ k = \lambda=1 \ , \ \eta=0 \ ,  \ p_\eta=0$. }
\end{figure}

\begin{figure}[ H]
\centering
\begin{subfigure}{0.4\textwidth}
\centering
    \includegraphics[width=\textwidth]{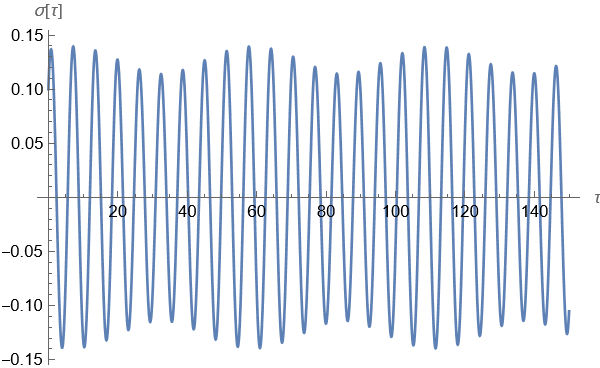}
    \caption{$\gamma=0.25$}
    \label{fig:11}
  \end{subfigure}
 \hfill
  \begin{subfigure}{0.4\textwidth}
  \centering
    \includegraphics[width=\textwidth]{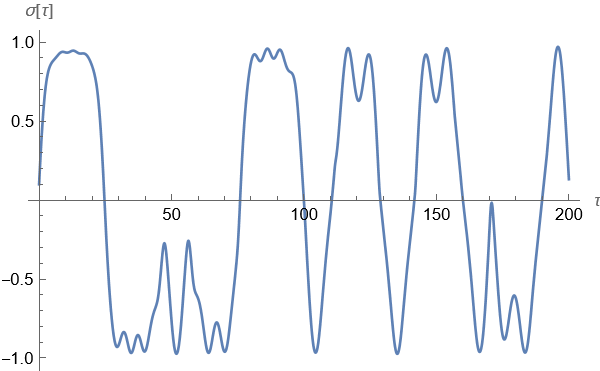}
    \caption{$\gamma=5$}
    \label{fig:22}
  \end{subfigure}
   \caption{Motion in ($\sigma-\tau$) plane for $\gamma$-deformed ST case with $E=2  \ , \ k= \lambda=1  \ , \ \eta=1 \ , \  p_\eta=0$. }
\end{figure}

We obtain distorted KAM tori and positive Lyapunov exponents for sufficiently large values of $\gamma$.
On the other hand, we observe an integrable dynamics for sufficiently small values of $\gamma << 1$.
This allows us to interpret the $\gamma$-deformation as an interpolation between an integrable and non-integrable dynamics sitting at two different extrema of the parameter space (see Fig.(\ref{fig:chaosST})). 
A careful analysis reveals that the string hits $\sigma \sim 1$ singularity \eqref{f:atd}, \eqref{f:natd} when the deformation parameter is large enough, $\gamma \gg1$.
One the other hand, the string never reaches the above singularity for small values of the deformation ($\gamma << 1$) parameter (see Figs.(\ref{fig:1})-(\ref{fig:2}) and (\ref{fig:11})-(\ref{fig:22})). 
From the bulk perspective, we therefore argue that the chaotic motion of these semi-classical strings could be an artefact of $\sigma \sim 1$ singularity as seen by the ``extended" string for large values of the deformation parameter ($\gamma >>1$).

From Figs.(\ref{fig:1})-(\ref{fig:2}) and (\ref{fig:11})-(\ref{fig:22}), one could see that the string rapidly touches $\sigma \sim 1$ singularity as the deformation is increased. This results in a sudden change in the string embedding which eventually causes a chaotic motion.
To summarise, one must therefore treat the $\gamma$-deformed solution \eqref{dgm} carefully at large values of the deformation parameter ($\gamma$) and see whether it is a trustable classical string background beyond certain limit \cite{Lunin:2005jy}.

From the perspective of the dual $\mathcal N =1$ SCFTs, this could be an artefact of the fragmentation of the dual operator \cite{Roychowdhury:2021eas}. 
One could imagine, in the simplest possible scenario heavy single trace operators $\mathcal O \sim$ Tr $\Phi^{ij}$ ; where $\Phi^{ij}$s are adjoint matter of some vector multiplet $V_i$. 
For large $\gamma$-deformations these operators are broken and the associated spin-chain structure is lost. 
A detailed understanding of the underlying mechanism is still lacking at this moment and certainly deserves further investigations in the future.

\vskip .3in
 
\noindent {\bf\large Acknowledgement}

 \vskip .2in

\noindent

We are indebted to Carlos Nunez for clarifying several issues along with his key insights on various parts of our work. 
The authors J.P. and D.R. are indebted to the authorities of IIT Roorkee for their unconditional support towards researches in basic sciences. 
D.R. also acknowledges The Royal Society, UK for financial support.
%%%%%%%%%%%%%%%%%%%%%%%%%%%%%%%%%%%

\appendix

\section{String motion for zero and general windings} \label{ap A}

In this appendix, we provide plots for the Poincar\'{e} sections corresponding to the string configurations with zero ($k = \lambda = 0 $) as well as general values of the winding numbers ($\{k,\lambda\} >1$). As illustrative examples, we choose $k = \lambda = 3 $ and $ k = \lambda =5 $ for the latter cases.

%%%%
\begin{figure}[h!]
	\centering
	\includegraphics[width=0.55\textwidth]{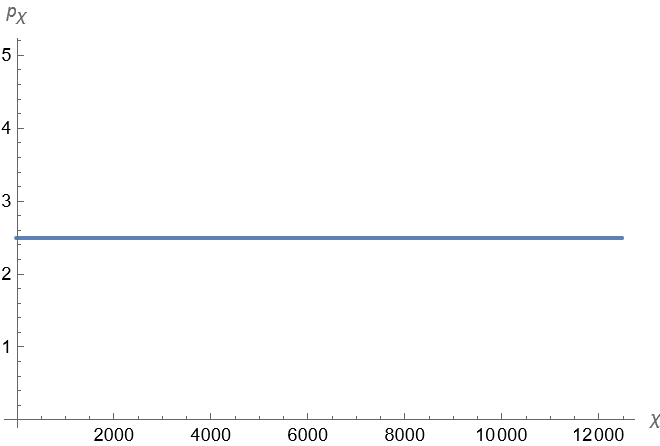}
	\caption{Poincar\'{e} section for point-particle limit ($k=\lambda=0$)
	corresponding to $\gamma$-deformed Abelian T-dual background.
	Here we fix the energy of the string as $E=6$ and the deformation
	parameters as $\gamma=0.25, 5$.}
	\label{fig:AP}
\end{figure}
%%%% 

%%%%
\begin{figure}[h!]
	\centering
	\begin{subfigure}[b]{0.4\textwidth}
		\centering
		\includegraphics[width=\textwidth]{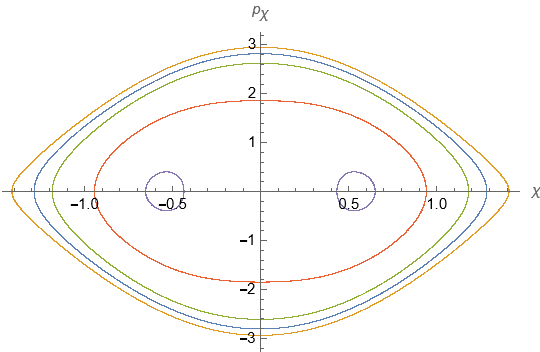}
		\caption{$E=6$, $\gamma=0.25$, $k= \lambda=3$ .}
		\label{fig:AA}
	\end{subfigure}
	\hfill
	\begin{subfigure}[b]{0.4\textwidth}
		\centering
		\includegraphics[width=\textwidth]{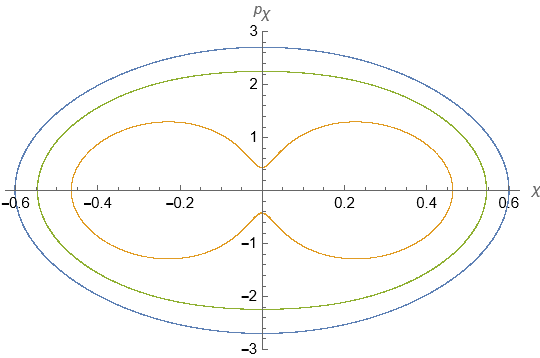}
			\caption{$E=6$, $\gamma=0.25$, $k= \lambda=5$ .}
		\label{fig:AB}
	\end{subfigure}
	\hfill
	\begin{subfigure}[b]{0.4\textwidth}
		\centering
		\includegraphics[width=\textwidth]{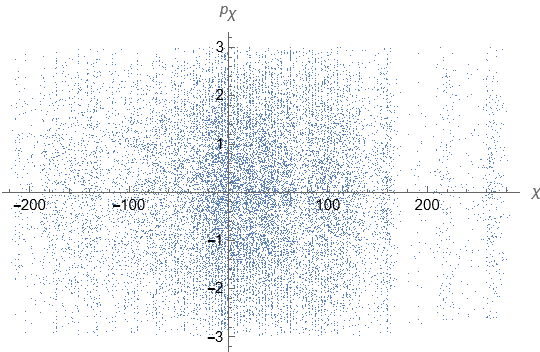}
		\caption{$E=6$, $\gamma=5$, $k =\lambda=3$ .}
		\label{fig:AC}
	\end{subfigure}
	\hfill
	\begin{subfigure}[b]{0.4\textwidth}
		\centering
		\includegraphics[width=\textwidth]{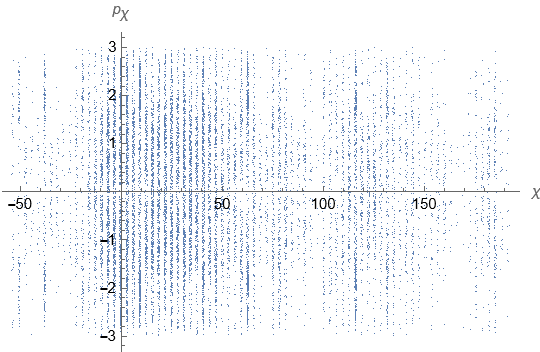}
		\caption{$E=6$, $\gamma=5$, $k=\lambda=5$ .}
		\label{fig:AD}
	\end{subfigure}  
	\caption{Poincar\'{e} sections for $\gamma$-deformed Abelian T-dual
	background with non-zero winding numbers. Here we fix the winding
	numbers as $k=\lambda=3$ in figs.(\ref{fig:AA}), (\ref{fig:AC})
	and $k=\lambda=5$ in figs.(\ref{fig:AB}),(\ref{fig:AD}). In all
	cases we set the energy $E=6$. The plots in the upper	row correspond to
	small deformation parameters ($\gamma =0.25$) whereas, those in the
	lower row correspond to large values of the deformation ($\gamma =5$).}
	\label{fig:AbPS}
\end{figure}
%%%%

%%%%
\begin{figure}[h!]
	\centering
	\includegraphics[width=0.55\textwidth]{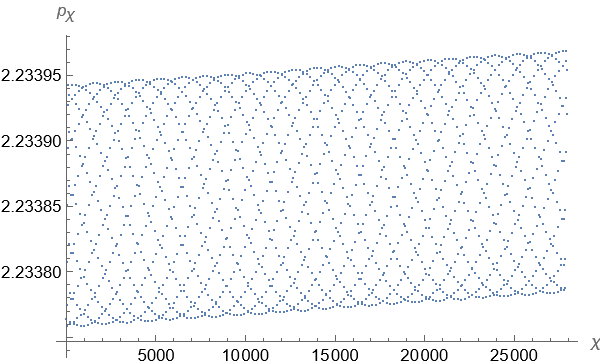}
	\caption{Poincar\'{e} section for point-particle limit ($k=\lambda=0$)
	corresponding to $\gamma$-deformed Sfetsos-Thompson background.
	Here we fix the energy of the string as $E=5$, and the deformation
	parameter as $\gamma=0.25$.}
	\label{fig:NFbSk1}
\end{figure}
%%%%

%%%%
\begin{figure}[h!]
	\centering
	\begin{subfigure}[b]{0.4\textwidth}
		\centering
		\includegraphics[width=\textwidth]{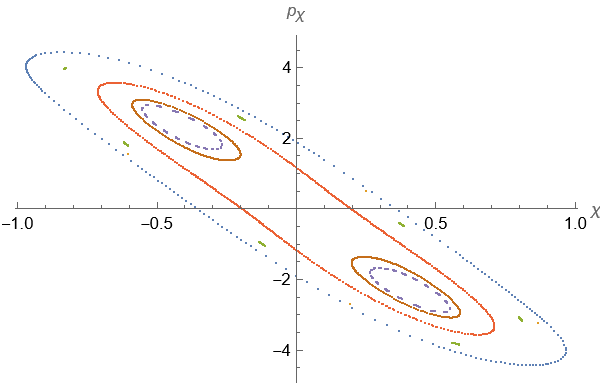}
		\caption{$E=5$, $\gamma=0.25$, $k=\lambda=3$ .}
		\label{fig:SA}
	\end{subfigure}
	\hfill
	\begin{subfigure}[b]{0.4\textwidth}
		\centering
		\includegraphics[width=\textwidth]{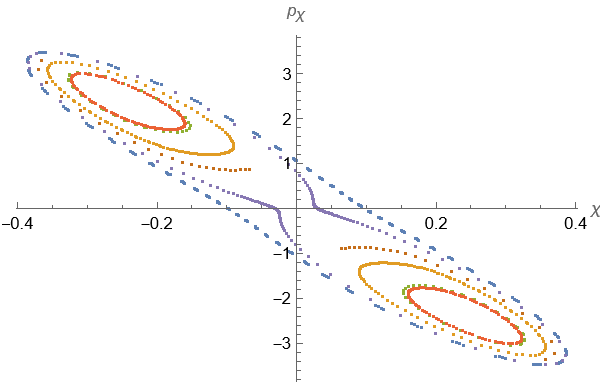}
		\caption{$E=5$, $\gamma=0.25$, $k =\lambda=5$ .}
		\label{fig:SB}
	\end{subfigure}
	\hfill
	\begin{subfigure}[b]{0.4\textwidth}
		\centering
		\includegraphics[width=\textwidth]{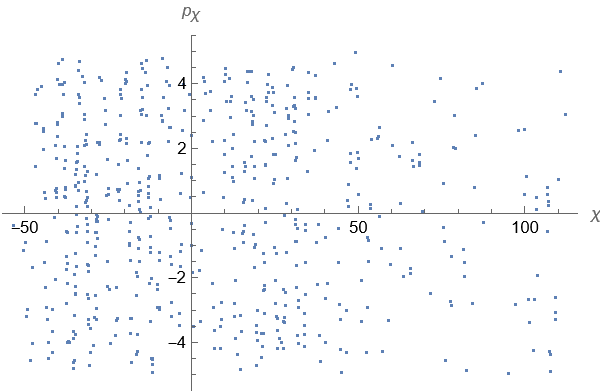}
		\caption{$E=5$, $\gamma=5$, $k= \lambda=3$ .}
		\label{fig:SC}
	\end{subfigure}
	\hfill
	\begin{subfigure}[b]{0.4\textwidth}
		\centering
		\includegraphics[width=\textwidth]{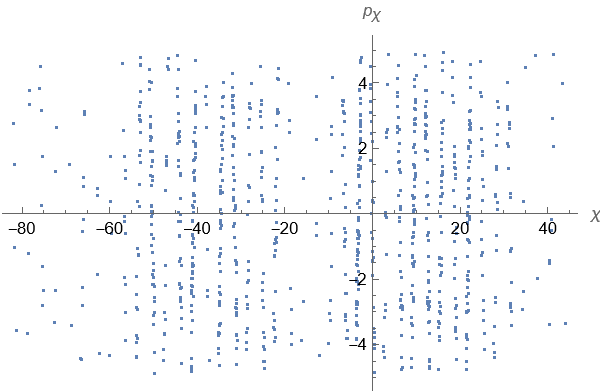}
		\caption{$E=5$, $\gamma=5$, $k= \lambda=5$ .}
		\label{fig:SD}
	\end{subfigure}  
	\caption{Poincar\'{e} sections for $\gamma$-deformed Sfetsos-Thompson
	background with non-zero winding numbers. Here we fix the winding
	numbers as $k=\lambda=3$ in figs.(\ref{fig:SA}), (\ref{fig:SC})
	and $k=\lambda=5$ in figs.(\ref{fig:SB}),(\ref{fig:SD}). In all
	cases we set the energy $E=5$. The plots in the upper	row correspond to
	small deformation parameters ($\gamma =0.25$) whereas, those in the
	lower row correspond to large values of the deformation ($\gamma =5$).}
	\label{fig:NAbPS}
\end{figure}
%%%%

As can be seen from fig.(\ref{fig:AP}), the point particle limits ($k=\lambda=0$) of the strings for the Abelian T-dual background preserve the integrability of the string $\sigma$-models irrespective of the values of the deformation parameters ($\gamma$). This is seen to follow from \eqref{EoM:HAb}. On the other hand, in the $\gamma$-deformed Sfetsos-Thompson case the point particle limits ($k=\lambda=0$) preserve integrability at small deformations only, as shown in fig.(\ref{fig:NFbSk1}). In the large deformation ($\gamma>>1$) the Hamilton's equation for $p_{\chi}$ (\ref{HNab:pchi}) can be expanded as
%%%%
\begin{equation}\label{pchi:lim}
\dot{p}_{\chi}=\frac{p_{\chi}^{2}\, \sin2\chi }{8\qty(\cos2\chi -1)^{2}
\sigma^{2}\gamma^{2}}+\mathcal{O}\qty(\gamma^{-3}) \, .
\end{equation}
%%%%

Clearly, in the limit $\chi\to 0$, $\dot{p}_{\chi}\to \infty$. This is the reason we do not get a phase space plot at large deformation. This also indicates the fact that we can not take arbitrarily large values of the deformation parameter ($\gamma$) \textendash{} we need to truncate $\gamma$ at some finite value to obtain reasonable solution. In this regard, the point particle limit corresponding to the $\gamma$-deformed Sfetsos-Thompson background may be considered as a special case.

Nevertheless, the (non-)integrability of the $\sigma$-models for (large) small values of the deformation parameter ($\gamma$) remain preserved even when we set the winding numbers of the strings greater than $1$ ($k=\lambda=3, 5$). These have been shown in figs.(\ref{fig:AbPS}), (\ref{fig:NAbPS}).

\section{Background dilaton and semi-classical analysis }

In $\gamma$-deformed Sfetsos-Thompson solution, for sufficiently small values of $\gamma$ ($\gamma <<1$) the string moves close to $\sigma \sim 0$ as described in Fig.(\ref{fig:11}). 
For $\gamma <<1$ and around $\sigma \sim 0$ the background dilaton in \eqref{dgmf} takes the form 
$$
e^{2\Phi}|_{\gamma <<1} \sim \text{const. + small fluctuations} \ ( \mathcal O(\gamma^2 \sigma^2)) \ . 
 $$
Hence for $\gamma <<1$, the $\alpha^\prime$ correction term ($ \alpha^\prime \sqrt{g} \mathcal R^{(2)} \Phi$) in the  {\it{Polyakov}} action
\eqref{action} gives a constant contribution (ignoring the small fluctuations) and the equations of motion still remain same.
Therefore, for $\gamma <<1$ the semi-classical analysis holds.

On the other hand, for sufficiently large values of $\gamma$ ($\gamma >> 1$) 
the background dilaton in \eqref{dgmf} takes the form 
$$
e^{2\Phi}|_{\gamma >> 1} \sim \frac{1}{\gamma^2 \sigma^2 \big(1-\sigma^2)^3} +  \mathcal O \Big(\frac{1}{\gamma^3}\Big) \ . 
 $$
As described in Fig.(\ref{fig:22}) we observe that the string rapidly touches $\sigma \sim 1$ for $\gamma >> 1$. 
Therefore, for $\gamma >> 1$ the background dilaton blows up and the semi-classical approximation breaks down. 
In other words, the $\alpha^\prime$ corrections are important in the limit of large $\gamma$ deformations. 
Similar remarks hold for the $\gamma$-deformed Abelian T-dual example which we therefore prefer not to repeat here.

\end{document}